\documentclass{nature}

\title{Intermediate states in Andreev bound state fusion}

\author{Christian J\"unger$^{1}$, Sebastian Lehmann$^2$, Kimberly A. Dick$^{2,3}$, Claes Thelander$^2$, Christian Sch\"onenberger$^{1,4}$ \& Andreas Baumgartner$^{1,4}$}

\usepackage{graphicx}
\makeatletter
\let\saved@includegraphics\includegraphics
\AtBeginDocument{\let\includegraphics\saved@includegraphics}
\renewenvironment*{figure}{\@float{figure}}{\end@float}
\makeatother

\usepackage{float}
\restylefloat{table}

\begin{document}

\maketitle

\begin{affiliations}
 \item Department of Physics, University of Basel, Klingelbergstrasse 82, CH-4056 Basel, Switzerland
 \item Division of Solid State Physics and NanoLund, Lund University, S-221 00 Lund, Sweden
 \item Center for Analysis and Synthesis, Lund University, S-221 00 Lund, Sweden
 \item Swiss Nanoscience Institute, University of Basel, Klingelbergstrasse 82, CH-4056, Basel, Switzerland
\end{affiliations}
%
%
%
\begin{abstract}
Hybridization is a very fundamental quantum mechanical phenomenon, with the text book example of binding two hydrogen atoms in a hydrogen molecule. In semiconductor physics, a quantum dot (QD) can be considered as an artificial atom, with two coupled QDs forming a molecular state,\cite{Wiel2002} and two electrons on a single QD the equivalent of a helium atom.\cite{Ellenberger2006}
Here we report tunnel spectroscopy experiments illustrating the hybridisation of another type of discrete quantum states, namely of superconducting subgap states\cite{Prada2020} that form in segments of a semiconducting nanowire in contact with superconducting reservoirs. We show and explain a collection of intermediate states found in the process of merging individual bound states, hybridizing with a central QD and eventually coherently linking the reservoirs. These results may serve as a guide in future Majorana fusion experiments and explain a large variety of recent bound state experiments.
\end{abstract}
%
%
\begin{figure}[th]
	\begin{center}
	\includegraphics[width=0.5\columnwidth]{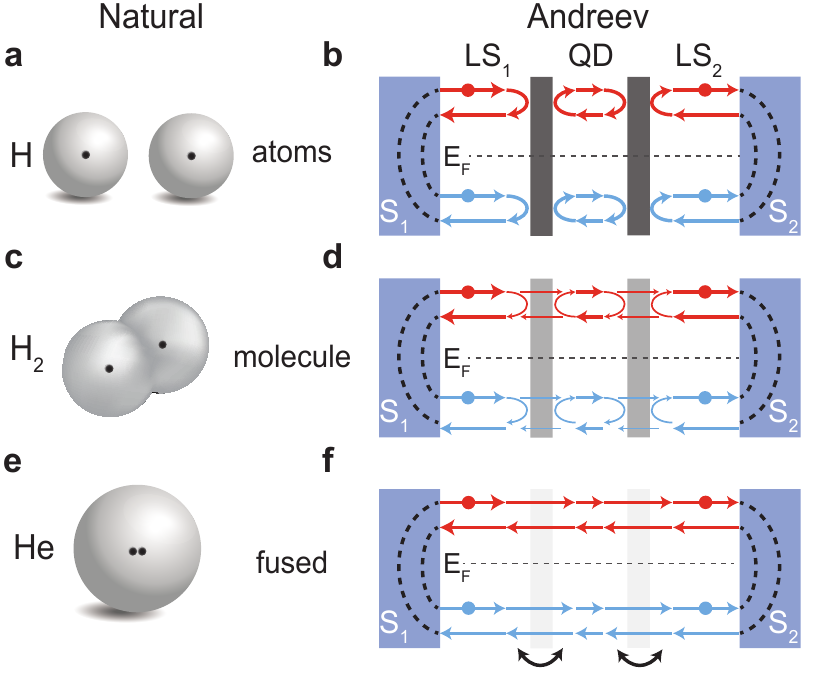}
    \end{center} 	
	\caption{\textbf{Atomic and Andreev state fusion.} \textbf{a} Comparison of the hybridization of natural atomic states (left) and Andreev bound states (right). The ABSs form in the two lead segments LS$_{1,2}$ coupled to the superconducting reservoirs S$_{1,2}$ with Andreev reflections at the S - LS interfaces. A central QD serves as a tunable coupler element. \textbf{a} and \textbf{b} show atomic, \textbf{c} and \textbf{d} molecular and \textbf{e} and \textbf{f} fused or helium states.}
	\label{fig:fig1}
\end{figure}
\begin{figure}[b]
	\begin{center}
	\includegraphics[width=0.5\columnwidth]{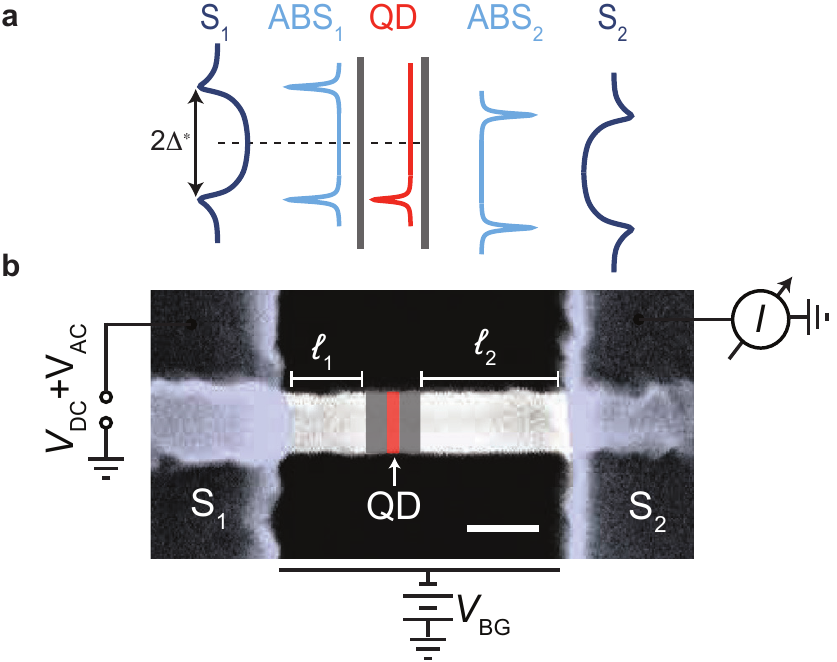}
    \end{center} 	
	\caption{\textbf{Energy level diagram and device.} \textbf{a} Schematic energy level diagram for the three NW segments corresponding to the scanning electron micrograph in \textbf{b}. In the LSs and on the QD the confinement results in discrete electronic states of various types. \textbf{b} Scanning electron micrograph of a representative device, including the measurements setup. The device consists of a zincblende InAs NW with an integrated QD (red, $\sim 25\,$nm), confined by two $30\,$nm wurtzite segments (gray) forming tunnel barriers due to a conduction band offset of $\sim$ \SI{100}{\milli eV} between the two crystal phases.\cite{Lehmann2013, Nilsson2016, Chen2017,Juenger2019} The superconducting contacts S$_{1,2}$ (blue) are coupled to the NW lead segments LS$_{1,2}$, one of which is shorter than the other, $\ell_1$ $<\ell_2$. The scalebar is $100\,$nm. In the experiments, we measure the modulation of the electrical current $I$ at the grounded drain terminal (S$_2$) and record the differential conductance $G=dI/dV_{\rm SD}=I_{\rm ac}/V_{\rm ac}$ as a function of the backgate voltage $V_{\rm BG}$ and the bias voltage $V_{\rm SD}$ applied to the source contact (S$_1$).}
	\label{fig:fig2}
\end{figure}

%
%
%
%
\newpage
For non-interacting confined electrons, single particle quantum states can be viewed as standing waves at discrete energies. Similarly, with a superconductor as a boundary, two-particle standing waves can form based on Andreev reflection (AR), as illustrated in Fig.~\ref{fig:fig1}.
Such discrete Andreev type bound states (ABSs), or `Andreev atoms', can form in a semiconductor with one,\cite{Lee2013,Gramich2017} two,\cite{Pillet2010,Prada2020} or potentially with multiple superconducting contacts.\cite{Riwar_Jouzet_Meyet_Nazarov_NatureComm_2016} We also use the term ABS for the conceptually similar Yu-Shiba-Rusinov (YSR) subgap states forming due to the interaction with the quasi-particles of the superconductor,\cite{GroveRasmussen2018, Prada2020,Scheruebl2020} and specify only when necessary. ABSs are the topologically trivial next of kin to the topologically non-trivial Majorana bound states (MBSs), currently a major research topic due to the promise of topologically protected quantum information processing.\cite{OFarrell2018} Merging MBSs in a process called `fusion' will be a first step in probing the nature of MBSs.\cite{Alicea2011,Zhang2019} First experimental signatures of MBSs were reported in recent years,\cite{Mourik2012,Albrecht2016, Nichele2020,Razmadze2020} but a controlled fusion experiment is hampered, for example, by the lack of sharp, gate tunable tunnel barriers and correspondingly well-defined NW segments.
To obtain such well-defined structures, we have grown InAs nanowires (NWs) with integrated tunnel barriers forming a central quantum dot (QD),\cite{Lehmann2013,Nilsson2016} and fabricated superconducting contacts S$_{1,2}$ at each end of the NW lead segments (LSs) adjacent to the QD. The ABSs form in these LSs.\cite{Juenger2019,Juenger2020} An energy level diagram and a colored scanning electron micrograph of such a device is shown in Fig.~\ref{fig:fig2}, including the experimental setup. The hybridization process is illustrated in Fig.~\ref{fig:fig1}: an increase in the gate voltage increases the tunnel coupling,\cite{Juenger2019,Thomas2020} so that the ABSs hybridize with the QD, forming one of the possible `Andreev molecules'.\cite{Su2017,Scheruebl2019,Pillet2019} For even lower barriers, ABSs extend over the complete device, eventually generating the Josephson effect. In analogy to the fusion of hydrogen atoms, one might call this state `Andreev helium'.
%
%
In all presented experiments we measure the differential conductance $G$, as discussed in Fig.~S2 and in the Methods section. The basic characterization of the central QD with the contacts in the normal state ($B=50\,$mT) is shown in Fig.~S1, with addition energies around \SI{8}{\milli eV}, a level spacing of \SIrange{0.5}{2}{\milli eV}, and a resonance broadening of $\Gamma_{\rm QD}\approx 50\,\mu$eV at low gate voltages, all increasing with increasing gate voltage.\cite{Nilsson2016, Juenger2019, Thomas2020}

\clearpage

\begin{figure}[h!]
	\centering
	\includegraphics[width=16.6cm,keepaspectratio]{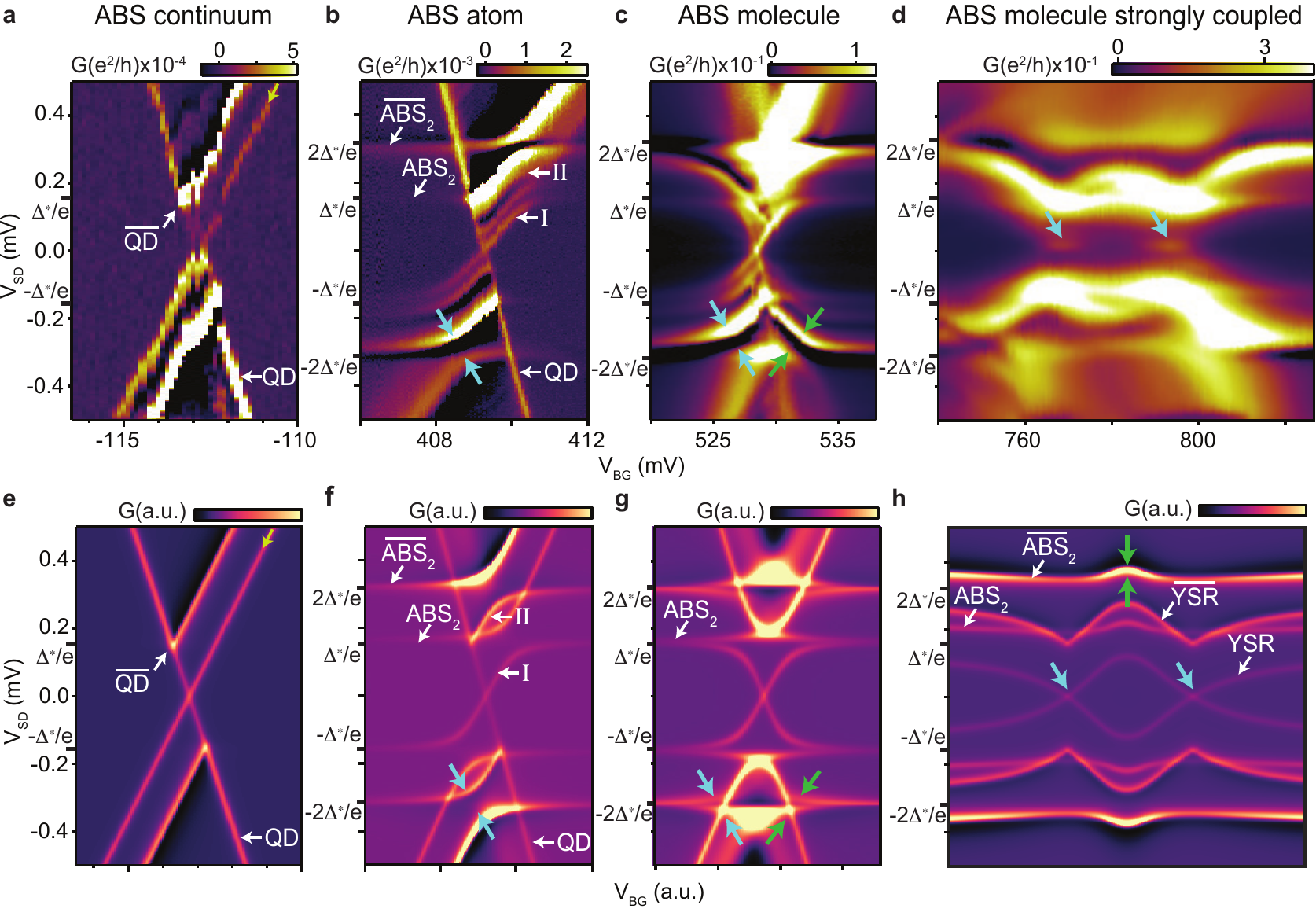}
	\caption{\textbf{Andreev atom, molecule and state fusion} Top row: $G$ vs. $V_{\text{BG}}$ and $V_{\text{SD}}$ in the superconducting state ($B=0$) for different gate voltage intervals, showing the intermediate states in the fusion process of ABSs in the three NW segments. Bottom row: results of the model discussed in the text. \textbf{a} and \textbf{e} Both LSs in the long junction limit with ABSs filling the energy gap, only showing the QD resonance. \textbf{b} and \textbf{f} Discrete ABS (`Andreev atom') in the shorter LS, showing an avoided crossing with the QD state. \textbf{c}  and \textbf{g} Discrete ABS in both LSs, both hybridizing with the QD state. \textbf{d} and \textbf{h} ABS in both LSs tunnel coupled to a YSR state on the QD (`Andreev molecular state'). The cyan arrows point out the onset of a Josephson current, carried between the superconducting reservoirs by the fused ABSs. The model parameters are given in table I of the Supplemental Material.}
	\label{fig:fig3}
\end{figure}

%
%
%
%
%
%
%
%
%
Our main results are plotted in Fig.~\ref{fig:fig3}, where $G$ is plotted as a function of the bias and gate voltage around four different charge degeneracy points of the QD, with the reservoirs in the superconducting state ($B=0$). Larger scale data and similar resonances in between are shown in Figs.~S2 and S3 respectively. The spectra at low backgate voltages, shown in Fig.~\ref{fig:fig3}a, exhibit Coulomb blockade (CB) resonances similar as in the normal state. In addition, we find a replica of the QD resonance of positive slope labeled $\overline{\rm QD}$, consistent with a QD state aligning with an increased quasi particle density of states (DoS) at an energy $\Delta^{*}\approx 160\,\mu$eV in LS$_1$ or in S$_1$.\cite{Gramich2015,Juenger2019}
For a QD directly, but weakly tunnel coupled to {\it two} fully gaped superconducting reservoirs, one would expect a transport gap for $e|\Delta V_{SD}| <  2\Delta^{*}$ and no shift of the CB diamonds in gate voltage.\cite{Gramich2015} Our finding can be understood by noting that the Fermi velocity in the NW increases with the gate voltage, so that the ABSs in the shorter lead segment, LS$_1$, is in the short-junction limit, with ABS energies only near $\Delta^{*}$, while the ABSs in the longer LS are still in the long-junction limit, with ABS energies distributed and broadened across the energy gap.\cite{Juenger2019} The subgap resonances we attribute to a residual single particle DoS in S$_1$ or in LS1,\cite{Su2018,Juenger2019} or to resonant Cooper pair tunneling,\cite{Gramich2015} both mapping at least parts of the normal state resonance.\par
\clearpage
%
%
At slightly larger gate voltages, the spectrum in Fig.~\ref{fig:fig3}b shows discrete, weak ABS resonances at constant $V_{\rm SD}\approx \pm \Delta^{*}/e$ (ABS$_2$), and stronger ones at $V_{\rm SD}=\pm 2\Delta^{*}/e$ labeled $\overline{\rm ABS}_2$. Larger scale data are shown in Fig.~S2.
The ABSs in the LSs becoming discrete with increasing gate voltage is consistent with an increase in the Fermi velocity and a correspondingly larger level spacings.\cite{Juenger2019,Thomas2020} The ABSs in the LSs are pinned to the source or the drain electrochemical potentials, $\mu_{\rm 1,2}$, due to the respective AR processes. In Fig.~\ref{fig:fig3}b, we find resonances at $V_{\rm SD}=\Delta^{*}/e$, suggesting that each LS is now in the short junction limit, with a small residual single particle DoS, or slightly overlapping ABSs. These discrete ABSs can be viewed as `Andreev atoms'. 
Most prominently, Fig.~\ref{fig:fig3}b shows a strong avoided crossing of the horizontal ABS resonance, $\overline{\rm ABS}_2$, with the CB resonance of positive slope, indicated by cyan arrows. The avoided crossing can be understood as the hybridization of an ABS in LS$_2$ with a QD state, thus forming the equivalent of a `heteronuclear molecular state'. This state is mapped by the superconducting coherence peaks in S$_1$, or by an ABS in LS$_1$. Since each ABS is pinned to the respective reservoir, the ABS of the avoided crossings in Fig.~\ref{fig:fig3}b is necessarily located in LS$_2$. An avoided crossing suggests an increased coupling between the LSs and the QD, consistent with an increase in the normal state broadening, from $\Gamma_{\rm QD}\approx$ \SI{50}{\micro eV} for Fig.~\ref{fig:fig3}a to $\Gamma_{\rm QD}\approx$ \SI{150}{\micro eV} for Fig.~\ref{fig:fig3}b.\cite{Juenger2019, Thomas2020}. The negative differential conductance near the ABS resonances and the avoided crossings are due to the non-monotonic DoS in the superconductor or in the LSs with ABSs.
Two more resonances, labeled I and II, follow the curvature of the avoided crossing discussed above. Intuitively, I corresponds to aligning the hybridized ABS-QD state to a residual single particle DoS in S$_1$, or LS$_1$. The slightly shifted resonance II we tentatively attribute to an additional ABS in LS$_2$.
In Fig.~\ref{fig:fig3}c, at slightly larger backgate voltage, we find such avoided crossings almost symmetrically around the origin of the original CB diamond, as pointed out by green and cyan arrows at negative bias. Following a similar argument as above, this corresponds to discrete ABSs forming on both LSs, each hybridizing with the QD state when the energies align, i.e. {\it not} at the same gate voltage and bias values. Now the CB diamonds at opposite bias are not shifted with respect to each other, and both resonances crossing zero bias replicate the avoided crossings, consistent with replicas due to a background DoS.
Data over two QD charge degeneracy points at even higher backgate voltages are shown in Fig.~\ref{fig:fig3}d, for which we extract a QD broadening in the normal state of $\Gamma_{QD} \approx $~\SI{400}{\micro eV}$>\Delta^{*}$. Here, we find two intense and strongly broadened resonances, essentially mirror symmetric for negative biases. These resonances oscillate roughly between the biases $\Delta^*/e$ and $2\Delta^*/e$ when increasing the gate voltage, opening a gap in the transport characteristics, reminiscent of a gate-tunable ABS mapped by superconducting reservoirs.\cite{Pillet2010}
Due to the reduced barrier height, one might expect that the ABSs in the LSs hybridize and form another type of molecular Andreev state hybridizing all three sites, as illustrated in Fig.~\ref{fig:fig1}d.
We further point out two small conductance peaks at zero bias (cyan arrows), which we tentatively attribute to the onset of the Josephson effect, since similar structures were found in other NW Josephson junctions,\cite{Saldana_Grove_Nygard_PRL121_2018, Razmadze2020} and no Kondo effect was found in the normal state (not shown). A Josephson effect can be understood as ABSs `fusing' across the complete NW length, forming an ABS spanning across all three sites, as illustrated in Fig.~\ref{fig:fig1}f, and carrying coherent Cooper pairs. These peaks correspond to Josephson currents of a few ten fA, too small to be investigated in current bias experiments in our setup. In Fig.~S4 we show bias-spectroscopy data for $V_{\rm BG}$ between $1\,$V and $6\,$V, where the the subgap conductance becomes larger than above the gap, due to Andreev reflection through low barriers, while Fig.~S5 shows current bias experiments for $V_{\rm BG}$ between $10\,$V and $16\,$V, where we find an increasing, gate voltage modulated Josephson current,\cite{Doh2005,Nilsson2011} up to $\sim 0.5\,$nA. The latter value is small, but compatible with the large distance and tunnel barriers between the reservoirs. \par
To illustrate the formation of intermediate states in the ABS fusion process, we use a basic three-site model, with ABS resonances in each LS, a single QD state, and the tunnel coupling in between as perturbations. The conductance we obtain as the product of the transmissions through the individual sites. The main results are plotted in Figs.~~\ref{fig:fig3}e-h, while the model is described in more detail in the Supplemental Material.
First, we use a constant transmission in the LSs and a BCS DoS with a constant background in S$_1$ and a constant DoS in the S$_2$, mimicking the long junction limit in LS$_2$.\cite{Juenger2019} The results in Fig.~\ref{fig:fig3}e reproduce the data in Fig.~\ref{fig:fig3}a, including a replica $\overline{\rm QD}$ of the QD resonance due to the BCS coherence peaks in S$_1$.
Next, Fig.~\ref{fig:fig3}f shows the case of one discrete ABS in LS$_2$, tunnel coupled to the QD, and both reservoirs in the superconducting state. As discussed for Fig.~\ref{fig:fig3}b, we find horizontal resonances ($\overline{\rm ABS}_2$) and an avoided crossing at $\pm 2\Delta^{*}/e$, with replicas (ABS$_2$) due to a background transmission in the LSs and a residual DoS in S$_1$. Similarly, adding a discrete ABS in LS$_{1}$, we obtain Fig.~\ref{fig:fig3}g, with a qualitatively similar pattern as in Fig.~\ref{fig:fig3}c, especially the additional avoided crossings with the QD resonance of negative slope. 
Increasing the tunnel coupling strengths between the sites does not reproduce the experiments in Fig.~\ref{fig:fig3}d. However, if we insert a gate dependent YSR state on the QD, we find the spectrum plotted in Fig.~\ref{fig:fig3}h, which qualitatively reproduces the experiments, especially the two-lobed structure between $\Delta^*/e$ and $2\Delta^*/e$. We added a small background in the reservoirs to render the YSR resonances around zero bias visible. The YSR resonances obtain strong replicas ($\overline{\rm YSR}$) from the mapping with the BCS DoS, while the originally horizontal ABS$_2$ in LS$_2$ now exhibits avoided crossings with the YSR feature (green arrows). 
The onset of the Josephson effect we find roughly where the YSR resonances cross zero bias, since the NW transmission is at a maximum and the bias as a decoherence factor at a minimum. 
\par
In summary, we present transport spectroscopy measurements of intermediate states that occur when `fusing' individual ABSs in the leads of a NW coupled to a central QD, ultimately forming subgap states reaching across the entire NW. We show the evolution from Andreev atoms to different types of Andreev molecules and the fusion of individual states to `Andreev helium', ultimately carrying a Josephson current. We qualitatively reproduce the experimental data with a simple model, which supports our qualitative understanding. Surprisingly, the model requires a YSR state on the QD to reproduce the data, suggesting a proximity effect on the QD mediated by the ABSs in the LSs, which can be viewed as a physical implementation of the zero bandwidth model.\cite{GroveRasmussen2018} We expect that such transitions could be observed on the same CB resonance when controlling the LS and QD states individually by separate sidegates, which can, for example, be used to perform fusion experiments with topologically non-trivial subgap states, and later for braiding experiments.
%
%
\begin{methods}
%
%
%
%
The InAs nanowires were grown by metal-organic-vapor-phase epitaxy (MOVPE) and have an average diameter of \SI{70}{\nano\meter}. Two segments of wurtzite crystal phase (thickness: \SI{30}{\nano\meter}) are in-situ grown inside the NW. These segments act as atomically precise hard-wall tunnel barriers for electrons, because the ZB and WZ bandstructure align with a conduction band offset of $\sim$~\SI{100}{\milli eV} \cite{Nilsson2016,Chen2017}. The ZB segment in between, which defines the QD, has a width of \SI{25}{\nano\meter}.\\
The nanowires were transferred mechanically from the growth substrate to a degenerately p-doped silicon substrate with a $\text{SiO}_2$ capping layer (\SI{400}{\nano\meter}). The substrate is used as a global back gate. For the electron beam lithography we employ pre-defined markers and contact pads, 
The superconducting contacts consist of evaporated titanium/aluminium (Ti/Al: \SI{5}{nm}/\SI{80}{nm}). Before the evaporation step, the native oxide of the NW is removed by an Argon ion sputtering. The total length of the junction is $\approx$  \SI{450}{nm} and the QD is located closer to one of the superconducting contacts S$_1$ (L$_1 \sim$ \SI{50}{nm}) than to the other superconducting contact S$_2$ (L$_2 \sim $ \SI{320}{nm}). All measurements were carried out in a dilution refrigerator at a base temperature of \SI{20}{\milli\kelvin}. Differential conductance has been measured using standard lock-in techniques with a bias voltage modulation of $V_{\text{ac}} =$ \SI{10}{\micro V_{\rm rms}} at a frequency of $f_{\text{ac}} =$ \SI{278}{\hertz}).
%
%

\end{methods}

\begin{addendum}
 \item 
The authors thank R. Delagrange, D. Chevallier, C. Reeg, S. Hoffman, J. Klinovaja and D. Loss for the useful discussions and A. Mettenleiter for the help with the illustrations. This research was supported by the Swiss National Science Foundation by a) the project "Quantum Transport in Nanowires" granted to CS b) the National Center of Competence in Research Quantum Science and Technology (QSIT), and c) the QuantEra project SuperTop. We further acknowledge funding from the European Union's Horizon 2020 research and innovation program under grant agreement No 828948, project AndQC, the project QUSTEC, as well as the Swiss Nanoscience Institute (SNI). S.L., K.A.D. and C.T. acknowledge financial support by the Knut and Alice Wallenberg Foundation (KAW) and the Swedish Research Council (VR). All data in this publication are available in numerical form at \url{https://doi.org/10.5281/zenodo.5610019}.
\item[Correspondence] Correspondence should be addressed to A.B. (email: andreas.baumgartner@unibas.ch) or C.J. (email: christian.juenger@berkeley.edu).
 \item[Competing financial interests] The authors declare that they have no
competing financial interests.

\end{addendum}

\newpage

\section*{References}
\bibliography{literature}

\newpage

\section*{\large Supporting information to intermediate states in Andreev bound state fusion}

\author{Christian J\"unger$^{1}$, Sebastian Lehmann$^2$, Kimberly A. Dick$^{2,3}$, Claes Thelander$^2$, Christian Sch\"onenberger$^{1,4}$ \& Andreas Baumgartner$^{1,4}$}

\begin{affiliations}
 \item Department of Physics, University of Basel, Klingelbergstrasse 82, CH-4056 Basel, Switzerland
 \item Division of Solid State Physics and NanoLund, Lund University, S-221 00 Lund, Sweden
 \item Center for Analysis and Synthesis, Lund University, S-221 00 Lund, Sweden
 \item Swiss Nanoscience Institute, University of Basel, Klingelbergstrasse 82, CH-4056, Basel, Switzerland
\end{affiliations}

\renewcommand{\thefigure}{S\arabic{figure}}
\setcounter{figure}{0}

In this Supporting Information part we provide details on the three-site model, the model parameters used for the plots in Fig.~3e-h in the main text, and the supplementary data mentioned in the main text.
\newpage

\section*{Description of the three-site model}
To illustrate the formation of intermediate states in the ABS fusion process, we use a simple toy model with {\it three} sites, namely the two LSs (LS$_1$ and LS$_2)$ and the QD, with the tunnel couplings in between as a perturbation, resulting in the strongly simplified Hamiltonian
$$ H=H_{{\rm LS}_1}+H_{{\rm LS}_2}+H_{\rm d} + H_{{\rm LS}_1-{\rm d}}+H_{{\rm LS}_2-{\rm d}},$$
with $H_{{\rm LS}_j}=\sum_k \epsilon_{j,k}s_{j,k}^{\dagger} s_{j,k}$ the Hamiltonian and $s^{(\dagger)}_{j,k}$ the annihilation (creation) operator for two resonances, $k\in \{ 1,2 \}$, in the uncoupled LS$_j$. For simplicity, we assume identical bound states in the two LSs pinned to the electrochemical potential of the respective reservoir, $\epsilon_{j,k}=\mu_{j}+(-1)^k E_{\rm BS}$, with $E_{\rm BS}$ a constant bound state energy, i.e. with a negligible gate capacitance compared to the ones to source and drain. $H_{\rm d}=\epsilon_{d}d^{\dagger}d $ describes the single QD eigenstate at the energy $\epsilon_d=-e\alpha_{\rm{S}d}V_{\rm S}-e\alpha_{\rm{D}d}V_{\rm D}-e\alpha_{\rm{BG}d}V_{\rm BG}$, with the lever arms $\alpha_{\ell d}$  to the respective reservoir, $\ell \in \{{\rm S,D, BG}\}$, and $d^{(\dagger)}$ the annihilation (creation) operator on the QD. The hybridization between the sites we describe by the hopping terms between LS$_j$ and the QD, 
$H_{{\rm LS}_j-{\rm d}}=\sum_k \left[ t_{dj}(s_{jk}^{\dagger}d + c.c.)\right]$, with the hopping parameter $t_{dj}$, which we choose identically between all sites. More advanced theoretical treatments of multiple NW segments were published recently [1-3].

To roughly reproduce the experiments shown in Fig.~3d, we replace the single QD resonance above by {\it two} bias-symmetric subgap resonances on the QD segment, for which we chose the form of YSR states, [4] where $\epsilon_d$ of the QD enters the YSR resonance energy $E_{\rm YSR}=\pm \Delta^{*}\frac{1-g^2+w^2}{\sqrt{(1-g^2+w^2)^2 + 4*g^2}}$ using $g = \frac{4\pi D_N t^2}{U(1-x^2)}$, $w  = 0.5 gx$ and $x=1+2\epsilon_d/U$, with $t$ the tunnel coupling strength to the LSs, $U$ the QD charging energy and $D_N$ the normal state DoS.

To obtain the electrical current $I$, we simply assume an identical Lorentzian transmission $\mathcal{L}(E)$ for all cases, with identical and constant broadening for each eigenstate $n$, centered at the eigenenergy $E_{n}$, and with an amplitude $A_j^{(n)}$ given by the weights on site $j$, yielding the transmission through the three sites $T(E)=T_1T_dT_2$ with $T_j=\sum_n |A_j^{(n)}|^2\mathcal{L}(E-E_n)$, and the current [5,6]
\begin{eqnarray}\nonumber
I &=&\int_{-\infty}^{\infty} D_{1}(E+eV_{SD}) \cdot D_{2}(E) \cdot T(E)\\&& \bigl[f_{2}(E) - f_{1}(E+eV_{\rm SD})\bigr] dE,
\label{eq:RestunnelingSQDS}
\end{eqnarray}
with $f_{1,2}$ the Fermi distribution functions and $D_{1,2}$ the single particle DoS in the respective reservoirs. For the latter we use the standard BCS expression with a small broadening described by the Dynes parameter [7,8] and an additional constant background.

\clearpage
\newpage
\section*{Model parameters used in the main text}

\begin{table}[th!]
\begin{center}
    \begin{tabular}{| l | l | l |}
    \hline
    Symbol	 	   			& value 		   		& description \\ \hline \hline
		contacts:					&									&							\\ \hline \hline
    $\Delta^*$ 				& $160\,\mu$eV 		& energy gap in NW  \\ \hline
	  $\gamma$					& $10\,\mu$eV			& Dynes parameter (BCS broadening) \\ \hline
		$D_{\rm offset}$	& $0.5D_{\rm N}$	& Offset of BCS DoS in contacts \\ \hline
		$T_{\rm sample}$	& $50\,$mK				& sample temperature \\ \hline \hline
		
		lead segments:		&									&						\\ \hline
		$E_{\rm BS}$ 			& $160\,\mu$eV 		& bound state energy in LSs  \\ \hline
    $t_{jd}$					& $40\,\mu$eV			& hopping parameter \\ \hline
	  $T_{\rm offset}$	& 0.5							& background transmission in NW LSs \\ \hline \hline
		
		quantum dot:			&									&   \\ \hline
		$E_{\rm add}$			& $8.0\,$meV			& QD addition energy \\ \hline
		$\Gamma_1=\Gamma_2$&$5\,\mu$eV			& QD coupling (small for clarity) \\ \hline
		$\alpha_{1d}=\alpha_{2d}$& $0.4$		& QD lever arms of source and drain \\ \hline
		$\alpha_{\rm BG,d}$& $0.2$					& QD lever arm of backgate \\ \hline \hline
		
		YSR state:				&									&			\\ \hline
		$t$								& $3.0\,$meV			& coupling strength between QD and leads \\ \hline
		$D_{\rm N}$				& $6\cdot 10^{19}\,{\rm m}^{-3}$& normal state DoS in leads \\ \hline
		
    \hline
    \end{tabular}

\end{center}
\end{table}

We note that to obtain the result plotted in Fig.~3h, we had to reduce the lever arm of the central QD region by 80\%, suggesting that the bias drops partially at other sites, which makes microscopic modeling very difficult.


%
%

\section*{Additional data mentioned in the main text}

\begin{figure}[h]
	\begin{center}
	\includegraphics[width=0.99\textwidth,height=\textheight,keepaspectratio]{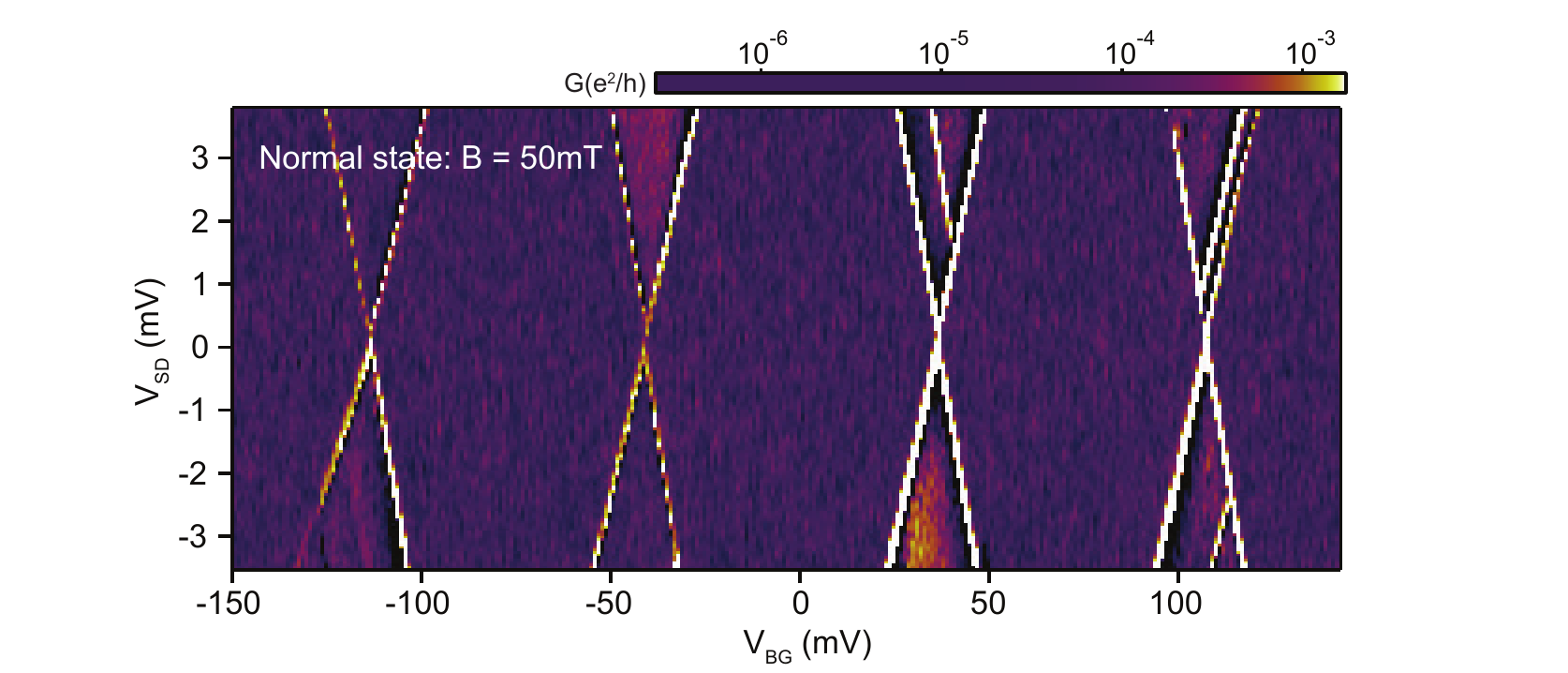}
    \end{center} 	
	\caption{\textbf{a} Normal state data at $B =\,$ \SI{50}{\milli \tesla}: Differential conductance $G$ plotted as a function of the backgate voltage $V_{\text{BG}}$ and the bias voltage $V_{\text{SD}}$. The QD is weakly coupled to the lead segments at these very low backgate voltages, with a line width of $\Gamma \approx$ \SI{50}{\micro eV}.}
	\label{fig:S1}
\end{figure}

\clearpage
%

%
\begin{figure}[t]
	\begin{center}
	\includegraphics[width=0.99\textwidth,height=\textheight,keepaspectratio]{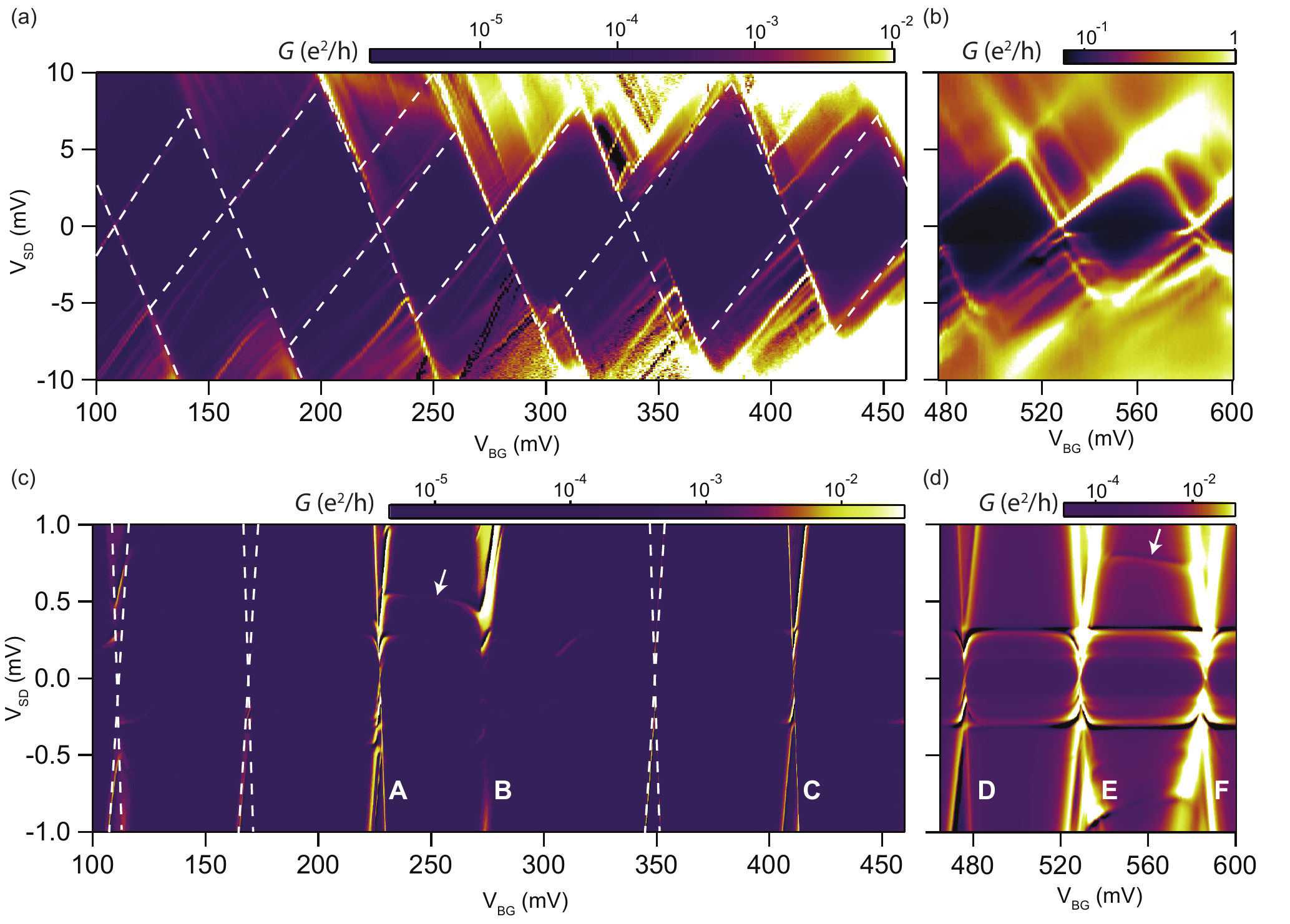}
    \end{center} 	
	\caption{\textbf{a} $G$ vs. $V_{\text{BG}}$ and $V_{\text{SD}}$ in the superconducting state ($B=0$) at intermediate gate voltages. (a) and (b) show an overview of CB diamonds, and (c) and (d) more detailed measurements at smaller biases. The high resolution data of resonances C and E are shown in main text in Figs.~3(b) and (c) and high resolution data of the resonances A, B, D, and F are shown in Fig. S3 below.}
	\label{fig:figS2}
\end{figure}

\clearpage
%

%
\begin{figure}[b]
	\begin{center}
	\includegraphics[width=0.99\textwidth,height=\textheight,keepaspectratio]{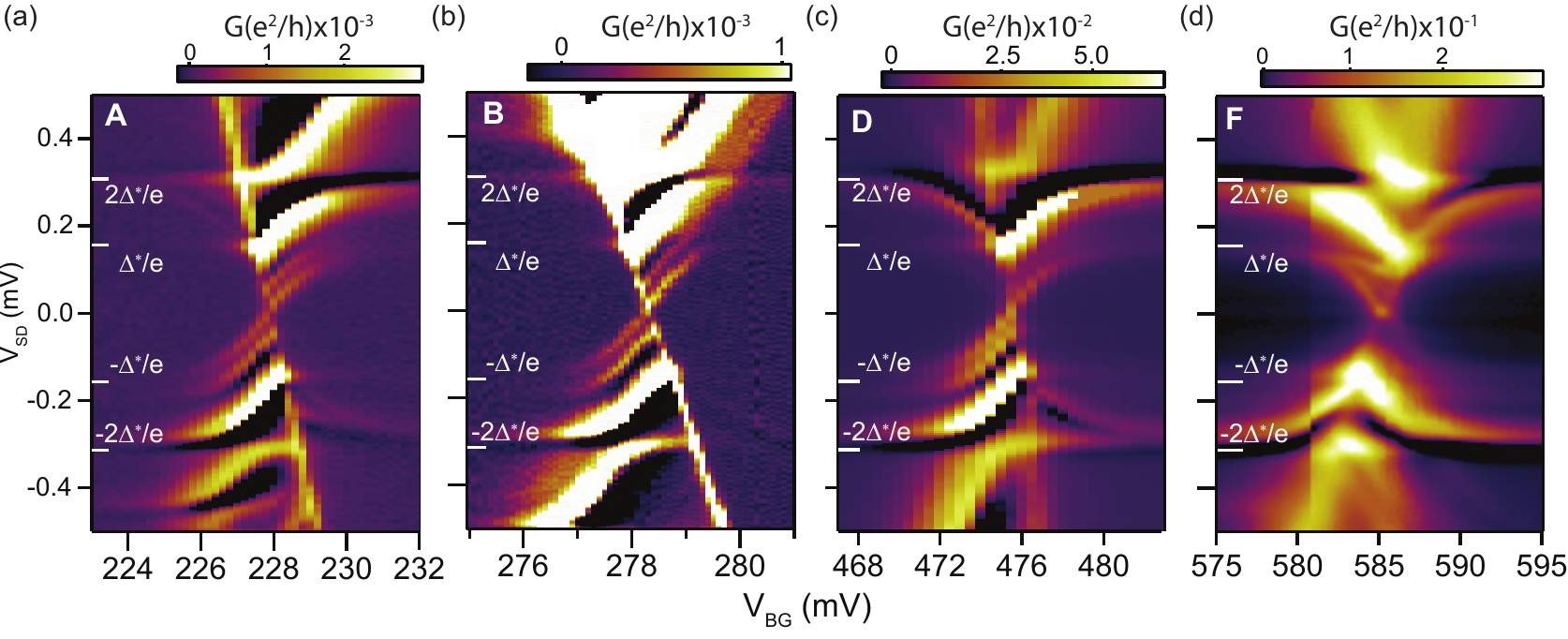}
    \end{center} 	
	\caption{\textbf{a} Detailed measurements of resonances A, B, D, F of Fig. S2, showing $G$ vs. $V_{\text{BG}}$ and $V_{\text{SD}}$ in the superconducting state ($B=0$).}
	\label{fig:figS3}
\end{figure}

\clearpage

%
%
\begin{figure}[t]
	\begin{center}
	\includegraphics[width=0.99\textwidth,height=\textheight,keepaspectratio]{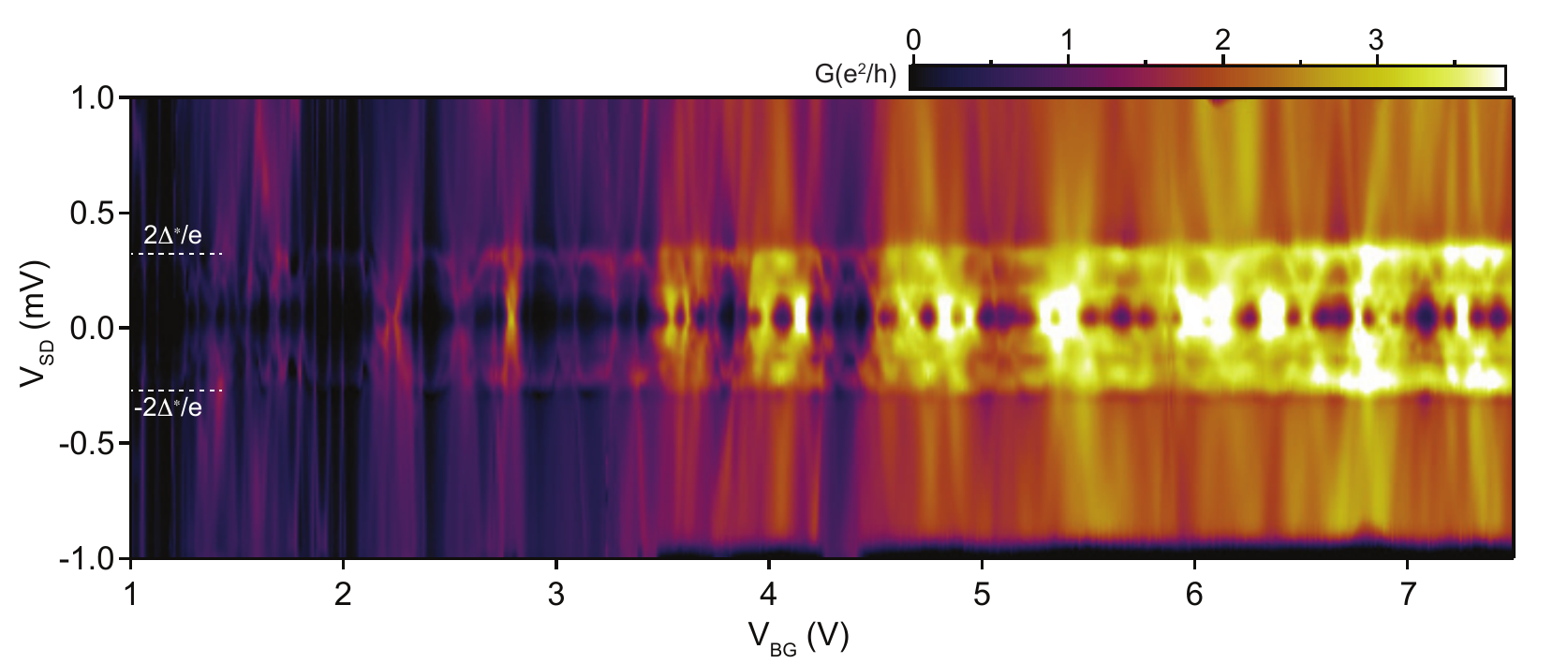}
    \end{center} 	
	\caption{\textbf{a} $G$ vs. $V_{\text{BG}}$ and $V_{\text{SD}}$ for larger backgate voltages and in the superconducting state ($B=0$), showing a consistently enhanced differential conductance for $|V_{\rm SD}|<2\Delta*/e$, consistent with a further increase in the tunnel coupling between the NW segments and a decreased barrier strengths..}
	\label{fig:figS4}
\end{figure}

\clearpage
%
%
%
\begin{figure}[b]
	\begin{center}
	\includegraphics[width=0.99\textwidth,height=\textheight,keepaspectratio]{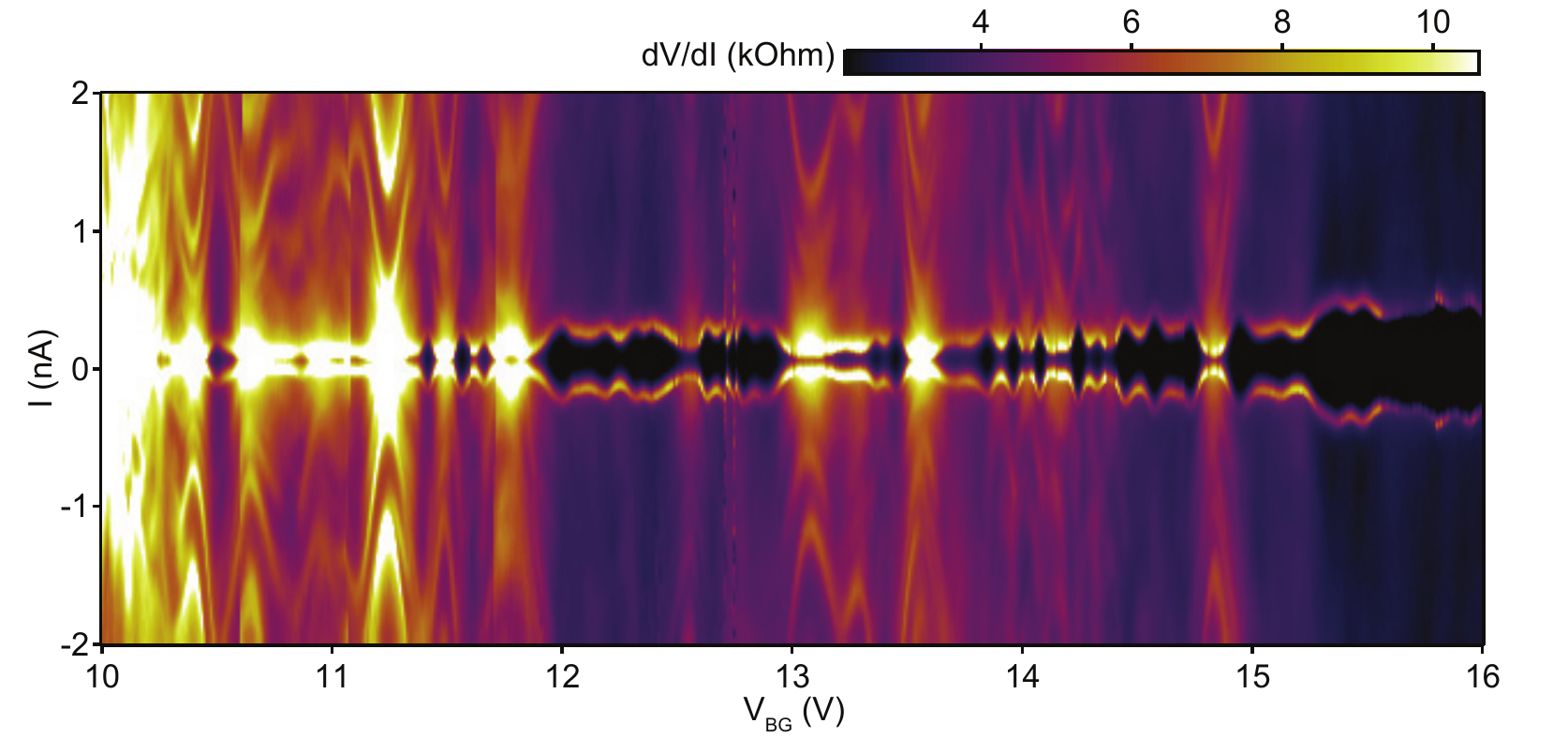}
    \end{center} 	
	\caption{\textbf{a} Current bias experiments and Josephson effect: here we plot the differential resistance $dV/dI$ as a function of $V_{\text{BG}}$ and the bias current $I$ in the superconducting state ($B=0$) for very large gate voltages, showing a gate volatge modulated zero-resistance Jospehson current and features due to multiple Andreev reflection (MAR), both suggesting a direct transmission of Cooper pairs between the superconducting reservoirs.}
	\label{fig:figS5}
\end{figure}

\clearpage

\section*{References}

\begin{enumerate}

\item Villas, A. \emph{et al.} Tunneling processes between yu-shiba-rusinov bound states, \emph{Phys. Rev. B} \textbf{103}, 155407 (2021).
  
\item Hess, R., Legg, H. F., Loss, D.,  Klinovaja, J.  Local and nonlocal quantum transport due to andreev bound states in finite rashba nanowires with superconducting and normal sections, \emph{Phys. Rev. B} \textbf{104}, 075405 (2021).

\item Steffenson, G.  O. \emph{et al.} Direct transport between superconducting subgap states in a double quantum dot. \emph{arXiv:2105.06815} (2021).

\item Kir\v{s}anskas, G.,  Goldstein, M., Flensberg, K., Glazman, L. I., Paaske, J. Yu-shiba-rusinov states in phase-biased superconductor–quantum dot–superconductor junctions. \emph{Phys. Rev. B} \textbf{92},  235422 (2015).

\item Yeyati, A.~L. ,  Cuevas, J.~C., L{\'{o}}pez-D{\'{a}}valos, A.,  Mart{\'{\i}}n-Rodero, A. Resonant tunneling through a small quantum dot coupled to superconducting leads. \emph{Phys. Rev. B} \textbf{55},  R6137– R6140 (1997).

\item Gramich, J.,  Baumgartner, A.,  Sch\"onenberger, C. Resonant and inelastic andreev tunneling observed on a carbon nanotube quantum dot. \emph{Phys. Rev.  Lett.} \textbf{115},  216801 (2015).

\item Dynes, R.~C., Narayanamurti, V.,  Garno, J.~P. Direct measurement of quasiparticle-lifetime broadening in a strong-coupled superconductor. \emph{Phys. Rev.  Lett.} \textbf{41},  1509 - 1512 (1978).

\item J\"unger, C. \emph{et al.} Spectroscopy of the superconducting proximity effect in nanowires using integrated quantum dots. \emph{Communications Physics} \textbf{2},  76 (2019).

\end{enumerate}

\end{document}